# Rare-earth-free noncollinear metallic ferrimagnets Mn$_{4-x}$Z$_x$N with compensation at room temperature

Rui Zhang[1], Yangkun He[1,*], Daniel Fruchart[2], J.M.D. Coey[1], Zsolt Gercsi[1]

1. School of Physics and CRANN, Trinity College, Dublin 2, Ireland
2. Institut Néel, CNRS & UGA, Dept QUEST, 38042, Grenoble Cedex 9, France

E-mail: heya@tcd.ie (Dr. Yangkun He)

**Abstract:**

Compensated ferrimagnets, like antiferromagnets, show no net magnetization but their transport and magneto-optic properties resemble those of ferromagnets, thereby creating opportunities for applications in high-frequency spintronics and low-energy loss communications. Here we study the modification the noncollinear ferrimagnetic spin structure of Mn$_4$N by a variety of metallic substitutions Z (Z = Cu — Ge and Ag — Sn) to achieve compensation at room temperature. The noncollinear frustrated 2.35μ$_B$ moments of Mn on 3$c$ sites of the (111) kagome planes tilt about 20° out-of-plane in Mn$_4$N and are easily influenced by the substitutions on 1$a$ sites, leading to different efficiency of compensation in Mn$_{4-x}$Z$_x$N that increases gradually from group 11 (Cu, Ag) to group 14 (Ge, Sn) with increasing number of valance electrons. Elements from the 5$^{th}$ period are more efficient to for compensation than those from the 4$^{th}$ period due to lattice expansion. The manganese site moments are determined by Z, orbital hybridization, charge transfer and the tilt angle, analyzed by constrained density functional theory. The Ga compound with compensation at room temperature for $x \approx 0.26$ is recommended for high-frequency spintronic applications.

**Keywords**: Metallic perovskites, Noncollinear magnetic structure, Kagome lattice, Ferrimagnetism, Compensation temperature, Mn$_4$N,

## 1. Introduction

Low energy loss communications and high-frequency spintronic applications could benefit from magnetic materials with a very low net magnetic moment at room temperature [1]. Although antiferromagnetic (AFM) materials have no net moment, apart from some exceptions with special symmetry [2,3], they usually lack net spin polarization and transport properties such as anomalous Hall effect. Ferrimagnetic (FiM) metals exhibit transport properties just like ferromagnets, while their magnetization can also fall to zero if the net moments of the antiparallel sublattices compensate, leading to the current interest in developing new compensated ferrimagnets for spintronics [4,5].

Ferrimagnets usually consist of two sublattices with different magnetic moments where AFM exchange may coexist with ferromagnetic (FM) interactions. The most common type, collinear ferrimagnetism, is shown in Fig. 1a; well-studied examples are Fe$_3$O$_4$, Y$_3$Fe$_5$O$_{12}$, amorphous Gd-Co [6] and Mn$_2$Sb [6,7]. In these materials, the moments within each sub-lattice are aligned parallel while the two sublattices are coupled antiferromagnetically. A less common non-collinear ferrimagnetic (ncFIM) magnetic structure can manifest, where the coupling is between a non-collinear antiferromagnetic sublattice, and a ferromagnetic sublattice as shown in Fig. 1b. The significant difference in the second type is that the ferrimagnetism is non-collinear. Examples include Ni(NO$_3$)$_2$ [8], MnCr$_2$O$_4$ [9] and Ho$_2$Fe$_{14}$B [10] at low temperature. Unlike oxides, which are usually insulating, metallic ferrimagnets are often R-T-based, where R is a heavy rare-earth and T is Fe or Co, or else Mn-based. The latter category avoids the use of rare-earth metals and often exhibits a high Curie temperature thus it is well suited for applications.

The metallic perovskite Mn$_4$N crystallizes in face-centred cubic (fcc) structure (space group Pm-3m) where N atoms occupying the body-centered interstitial site are coordinated by an octahedron six Mn atoms in the 3$c$ face-center positions, as shown in Fig. 2a. The Mn in the 1$a$ corner positions is not in direct contact with the nitrogen. The magnetic order has been investigated both experimentally and theoretically. Initially, it was thought to have a collinear ferrimagnetic structure along [111] [11,12] with a large 1$a$ – site magnetic moment $m_{1a}$ = 3.8 $\mu_B$ and a 3$c$ site moment $m_{3c}$ = 0.9 $\mu_B$ that was much-reduced by $p$-$d$ hybridization with the neighboring nitrogen [12]. Subsequent neutron diffraction with polarization analysis revealed that triangles of 3$c$ atoms in (111) planes, where they from a kagome lattice shown in Fig. 2b, add a triangular antiferromagnetic component to the 3$c$ sublattice spin structure, where the spins are either in the (111) plane where the spin axes either meet at the centre of the triangle ($\Gamma^{4g}$ mode) or else lie along the sides ($\Gamma^{5g}$ mode). Only a small anisotropy energy separates them, and the modes may coexist at finite temperature [13]. Calculations by Uhl et al. [14] confirmed a noncollinear 'umbrella' structure with the net sublattice moments aligned antiparallel along the [111] axis, producing a net moment close to the value of 1.1 $\mu_B$ that is found experimentally [12]. The $\Gamma^{4g}$ mode has a topological character that accounts for the large anomalous Hall effect in many Mn$_{4-x}$Z$_x$N materials [15]. The assumption that Mn$_4$N is a collinear ferrimagnet is therefore unwarranted and may lead to misleading conclusions [16].

In the Weiss mean-field model the temperature-dependent magnetization (*M-T*) curves for ferrimagnets can be categorised into three types depending on the moment and main molecular field coefficients within and between sublattices ($n_{1a1a}$, $n_{1a3c}$ and $n_{3c3c}$) [17]; curves are shown in Figs. 1c-1f. The *M-T* curve of binary Mn$_4$N [18,19] is *Q*-type, without compensation, where the magnetization of the 1$a$ sublattice is always larger than that of the 3$c$ sublattice and $n_{1a3c}$ is the main molecular field coefficient. An *N*-type *M-T* curve with compensation may be achieved by substitution Z for Mn on 1$a$ sites in Mn$_{4-x}$Z$_x$N (Z = Co, Ni, Cu, Zn, Ga, Ge, As, Rh, Pd, Ag, Cd, In, Sn, Sb, Pt, Au and Hg with x < 1) [20,21]. All of them are expected to produce compensation and some have been identified experimentally [22,23,24,25]. Though most dopants are nonmagnetic, it is found that compensation is achieved at quite different values of *x* for different elements [26]. Therefore, it is necessary to analyze the doping efficiency in order to clarify the governing physical mechanisms that allow us to productively design novel Mn$_4$N-based compensated ferrimagnets for room temperature applications.

In this study, we first reveal the origin of the noncollinear ferrimagnetism of Mn$_4$N. We then demonstrate compensation at room temperature with various non-magnetic alloying elements and compare their efficiency. We discuss the findings in relation to valance electron count, magnetic moment, tilt angle and lattice constant, based on the experimental data and constrained density functional theory calculations.

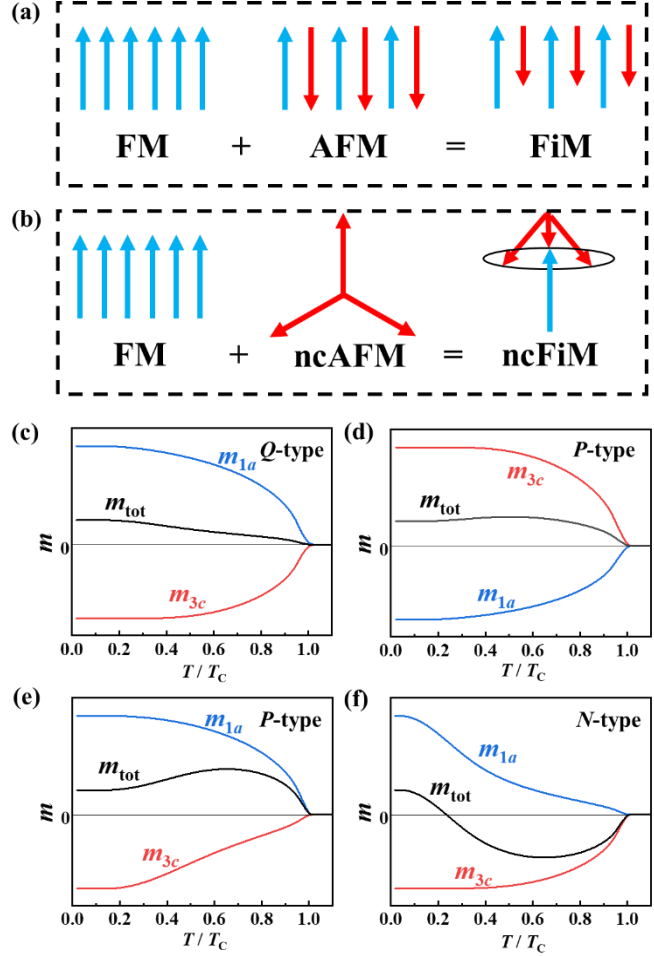

**Fig. 1. Ferrimagnetic prototypes.** (a) Collinear ferrimagnetic (FiM) spin structure is a combination of ferromagnetic (FM) and antiferromagnetic (AFM) spin structures. (b) Noncollinear ferrimagnetic (ncFiM) structure combines non-collinear antiferromagnetic (ncAFM) and FM structures. Three types of temperature-dependent magnetization curves (schematically) for a ferrimagnet with two sublattices ($1a$ and $3c$ for $Mn_4N$) with (c) $m_{1a} > 3m_{3c}$ with $n_{1a3c}$ as the main molecular field coefficient, (d) $m_{1a} < 3m_{3c}$ with $n_{1a3c}$ as the main molecular field coefficient, (e) $m_{1a} > 3m_{3c}$ with $n_{1a1a}$ as the main molecular field coefficient and (f) $m_{1a} > 3m_{3c}$ and $n_{3c3c}$ as the main molecular field coefficient. The collinear model is used here. In a noncollinear model the chief difference in the shape of the curve is the non-zero slope at low temperatures (see Fig. 2c).

## 2. Methods

The high purity (> 99.99%) elements of Mn and Z = Cu, Ga, Ge, In, Sn were arc-melted together five times to prepare homogeneous polycrystalline ingots. Additional Mn (2%) was added to compensate the loss due to its high vapor pressure. The ingots were then ground into powder and reacted with $N_2$ (> 99.99%) at 750 – 800 °C at a pressure of 50 kPa for 1 day. We found that if the $N_2$ pressure is too large (100 kPa) a $Mn_2N$ impurity phase will form in some samples with small values of $x$. Nitrogen deficiency can lead nitrogen vacancies or formation of γ- or β-Mn type impurities. Additional heat treatment (annealing at 660 °C in vacuum for one day was needed for Mn-Cu and Mn-Ge ingots before grinding them into powder to transform γ-Mn into β-Mn, owing to the ductile mechanical properties of γ-Mn which makes it difficult to grind.

The composition of the polycrystalline sample was checked by energy-dispersive X-ray spectroscopy. The crystal structure was characterized by powder X-ray diffraction (XRD) that showed a single-phase cubic structure. Magnetization measurements were conducted using a superconducting quantum interference device magnetometer (SQUID, Quantum Design).

*Ab*-initio calculations based on density functional theory were carried out using norm-conserving pseudopotentials and pseudo-atomic localized basis functions implemented in the OpenMX software package [27]. Calculations were based on a minimal 5 atom basis cell of the cubic structure using $13 \times 13 \times 13$ k-points to evaluate the total energies. Pre-generated fully relativistic pseudopotentials and the pseudo-atomic orbitals with a typical cut-off radius of 6 atomic units (a.u.) were used with $s3p3d3$ for the metal and $s3p3d2$ for the metalloid elements, respectively. A energy cut-off of 300 Ry was used for the numerical integrations. The convergence criterion for the energy minimization procedure was set to $10^{-8}$ Hartree. In the case of the non-collinear calculations, we show results without spin-orbit interaction (SOI), whose influence on the total energy is negligible compared with the exchange interaction

## 3. Results
### 3.1 Non-collinear ferrimagnetism in Mn$_4$N

The origin of non-collinear ferrimagnetism can be deduced from the magnetic interactions and the crystal structure, identified by X-ray diffraction in Fig. 2a. The lattice parameter $a_0 = 3.865$ Å is also the nearest-neighbour distance between two Mn$^{1a}$ atoms $d_{1a1a}$. The nearest distances between Mn$^{3c}$ and Mn$^{1a}$ or Mn$^{3c}$ $d_{1a3c}$ and $d_{3c3c}$ are both equal to $a_0/\sqrt{2} = 2.733$ Å. Generally, Mn atoms separated by 2.5-2.8 Å have delocalized electrons and couple antiferromagnetically while Mn atoms with longer separations (> 2.9 Å) couple ferromagnetically. Therefore, the Mn$^{1a}$ moments lie parallel to each other, whereas the small $d_{1a3c}$ distance favors antiparallel coupling between the sublattices. The separation of nearest-neighbor Mn$^{3c}$ atoms $d_{3c3c}$ is responsible for the non-collinear triangular antiferromagnetism of the 3$c$ sublattice. Together, these interactions lead to the umbrella-like spin structure, and the overall non-collinear ferrimagnetism.

Mn$_4$N has a high Curie temperature $T_C$ (780 K) and a small saturation moment $m_{tot} = 1.1$ μ$_B$/f.u. along a [111] direction, as shown in Fig. 2c. The measured moment $m_{tot}$ is the difference of the $m_{1a}$ and three times the ferrimagnetic component of Mn$^{3c}$ $m_c^{FiM}$, which are 3.8 μ$_B$ and -0.9 μ$_B$ per Mn, respectively [13]. It should be noted that the net moment in Fig. 2c remains a constant below 50 K and then drops with increasing temperature. By 160 K ($T/T_C = 0.2$), the moment has fallen by 13% of the 4 K value, in agreement with literature [18]. This is quite unusual, because according to the collinear mean-field model, the decrease at $T/T_C = 0.2$ should be smaller than 1% (see Fig. 1). The inability to fit the *P*-type curve to a collinear mean-field model for Mn$_4$N [19] is a strong indication of the noncollinear nature of the magnetic order.

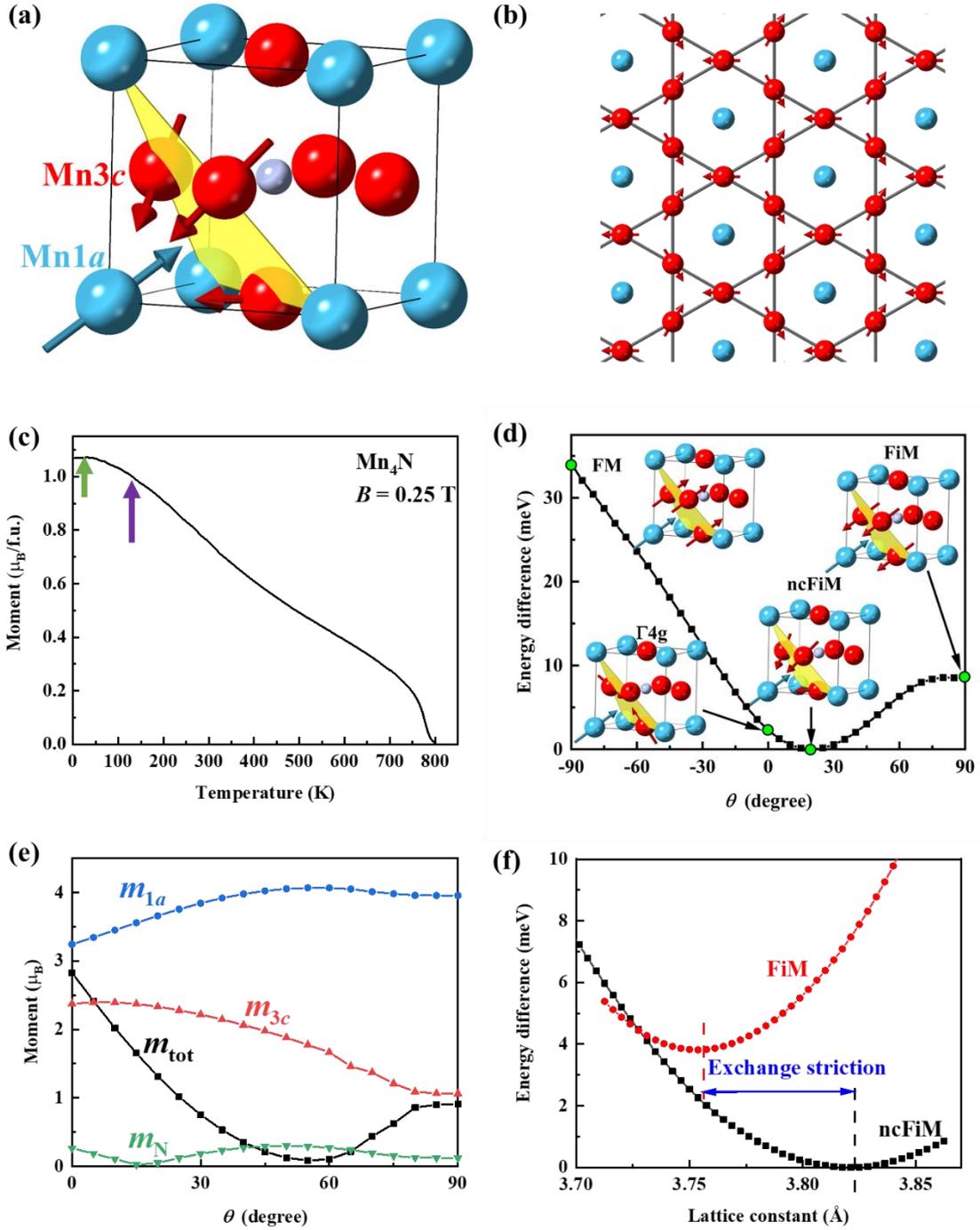

**Fig. 2. Noncollinear magnetic structure of Mn₄N.** (a) Crystal and magnetic structure showing the $\Gamma^{4g}$ triangular ferrimagnetism. Grey, blue and red atoms represent N, Mn$^{1a}$ and Mn$^{3c}$, respectively. (b) Kagome lattice of Mn$^{3c}$ in a (111) plane showing $\Gamma^{4g}$-like magnetic structure. The out-of-plane magnetic component is not shown. (c) Temperature-dependent magnetization $M(T)$ for Mn₄N. The magnetization remains constant up to 50 K (green arrow) but drops significantly above. At 160 K (purple arrow) the moment has already fallen to 87% of the base temperature value. (d) Energy difference in the calculated magnetic structure as a function of tilt angle $\theta$ between $m_{3c}$ and (111) plane. (e) Magnetic moments with varied tilt angle $\theta$. (f) Comparison of calculated total energies as a function of lattice constant for the collinear and noncollinear ferrimagnetic structures.

The noncollinear spin structure is analysed further using a constrained DFT approach, where the directions of the individual spins are pinned to a selected angle but the magnitudes of the moments are allowed to vary freely in the total energy minimization process. The direction of $Mn^{1a}$ is pinned to the body diagonal [111] and the tilt angle $\theta$ of the spins the $Mn^{3c}$ atoms is varied (also see inset of Fig. 5b). As the angle is rotated from the collinear ferrimagnetic configuration ($\theta = 90°$) into the (111) plane ($\theta = 0°$, $\Gamma^{4g}$ spin structure) and then towards a ferromagnetic configuration at $\theta = -90°$ the energy and magnetic moments vary as shown in Fig. 2d and 2e One can visualize this angle change like a closed umbrella (FiM) that opens out to close to 0° in regular usage and well beyond it on a windy day (FM, $\theta = -90°$). The relative total energy change of the constrained angle approach is shown Fig. 2d. Following the total energy minimum curve from the right to left, we witness a decrease in total energy $E_{tot}$ with the $Mn^{3c}$ moments canting away from the antiparallel spin arrangement into the (111) plane. The minimum of $E_{tot}$ is found at around $\theta = 20°$. The $E_{tot}$ difference between the collinear ferrimagnetic and noncollinear ferrimagnetic ground state is significant, about 4.5 meV/atom. The calculations also confirm that the FM arrangement ($\theta = -90°$) of the spins on Mn is very unfavourable with energy difference ~32 meV/atom. Our results suggest that the $Mn^{3c}$ sublattice moments make an angle close to 70° with the $Mn^{1a}$ moments, very far from the simplified picture of collinear ferrimagnetism often assumed. For further analysis, Fig. 2e shows the calculated site-specific magnetic moments together with the total magnetic moment per formula unit in our range of interest ($\theta = 0$ to 90°). A strong dependence of magnetization for both magnetic sites as a function of $\theta$ is revealed; the $Mn^{1a}$ moment remains ~4 $\mu_B$ down to about $\theta = 45°$ and then a significant reduction to ~3.2 $\mu_B$ occurs on closing towards the (111) plane ($\theta = 0°$). In contrast, the $Mn^{3c}$ site moment increases from about 1.1 $\mu_B$ up to 2.4 $\mu_B$ when the angle closes from FiM towards the $\Gamma^{4g}$-like configuration. The $E_{tot}$ minimum suggest magnetic values of $m_{1a}$=3.65 $\mu_B$, $m_{3c}$=2.35 $\mu_B$ with $m_{tot}$=1.24 $\mu_B$/f.u. Indeed, the collinear ferrimagnetic spin configuration also yields a value of $m_{tot}$ that is close to the experimental one, but we have relaxed both spin configuration for the equilibrium lattice parameters from DFT and find a value $a_0 = 3.75$ Å for the collinear ferrimagnetic state that is smaller than that for the non-collinear ferrimagnetic state $a_0 = 3.82$ Å, in Fig. 2f, which is closer to the experimental value of 3.865 Å at 300K. The earlier calculation [14] found a greater tilt angle and a smaller $3c$ moment, fixing the lattice parameter and exploring a smaller range of $\theta$. The energy difference can be ascribed to the electronic pressure caused by the altered magnetic spin configuration. This exchange striction, like that in FeRh [28], explains the significantly expanded lattice constant for $Mn_4N$ (3.86 Å) compared to its ferromagnetic cousins such as $Fe_4N$ (3.79 Å), $Co_4N$ (3.75 Å) and $Ni_4N$ (3.72 Å) on the one hand and on the other hand it is also manifest throughout the rotation of the spins that alters exchange-split band energies by Coulomb repulsion. This non-Heisenberg like behaviour relates to the spin split $d$-bands crossing the Fermi level that influences the band filling and calculated magnetic moments.

### 3.2 Doping for compensation

In order to achieve compensation, namely to change the temperature-dependent magnetization from $Q$-type to $N$-type, the main exchange should change from $n_{1a3c}$ to $n_{3c3c}$, meanwhile the $1a$ site moment $m_{1a}$ should be larger than three times the axial component of the $3c$ site moments $3m_{3c}^{FiM}$. This means that Mn on the $1a$ site should be substituted at the appropriate level $x$ in $Mn_{4-x}Z_xN$ (Z= Co, Ni, Cu, Zn, Ga, Ge, As, Rh, Pd, Ag, Cd, In, Sn, Sb, Pt, Au and Hg with $x < 1$). Fig. 3a shows the X-ray diffraction (XRD) pattern of $Mn_{3.76}Ga_{0.24}N$, and the low-angle data are expanded in Fig.

3b. The larger intensity of the (110) superlattice peak indicates that $Mn_{3.76}Ga_{0.24}N$ crystallizes in a well-ordered structure with Ga atoms occupying the 1$a$ site. Nonmagnetic Ga weakens the magnetic exchange leading to a decreased $T_C$ = 610 K. The net moment of 0.17 $\mu_B$/f.u. at 4 K indicates that each Ga decreases the moment by of ~3.8 $\mu_B$, matching both the moment of $m_{1a}$ from neutron diffraction [13] and our calculation. The compensation temperature is then $T_{comp}$ = 408 K. In this case, the 1$a$ sublattice dominates the magnetization at low temperatures, while the 3$c$ sublattice is dominant above compensation. The *M-H* curves shown in Fig. 3 exhibit very little hysteresis, indicating weak cubic magnetocrystalline anisotropy. Note the magnetization at 4 K is not saturated even in 5 T, further supporting the non-collinear ferrimagnetic structure where the tilt angle $\theta$ changes with magnetic field.

The doping efficiency of different elements from Cu to Sn is shown in Figs. 3e-3i. Unlike Ga, that changes the net moment at the rate of ~3.8 $\mu_B$/atom, the rates for the other elements are significantly different. For Sn with $x$ = 0.26, the magnetization is reduced to 0.26 $\mu_B$/f.u., much more than 0.12 $\mu_B$/f.u. for Ga with the same $x$. The magnetization curve shows a large hysteresis at 4 K in Fig. 3e, attributed to the $\Gamma^{5g}$ antiferromagnetic configuration that co-exist all the way down from the Néel temperature of $Mn_3SnN$ [29]. There is a difference in the *M-T* curves measured after field-cooling (FC) and zero-field-cooling (ZFC) shown in Fig. 3f, which is also observed in the In-doped sample ($x$ = 0.26). The compensation temperature is around 400 K for Sn substitution for $x$ = 0.26. The magnetization is less sensitive to $x$ for Sn than for Ga, and this trend towards low doping efficiency is more significant for Ge than Sn, as shown in Fig. 3g. Compensation was not observed below 400 K for Ge with $x$ = 0.35. On the other hand, $Mn_{4-x}Cu_xN$ is very sensitive to the compositional changes. The *M-T* curve gradually changes from *Q*-type to *N*-type and finally to *P*-type within a narrow range of $x$, as illustrated in Fig. 3h. The *M-T* curves for Cu, Ga, Ge, In and Sn with x = 0.26 are all compared in Fig. 3i). The Cu-doped sample has the highest doping efficiency with a *P*-type *M-T* curve without compensation ($T_{comp}$ < 0 K). Ge, Ga and Sn doped samples exhibit $T_{comp}$ of 70 K, 291 K and ~ 400 K respectively with *N*-type *M-T* curves. Ge leads either to a much higher $T_{comp}$ or else to a complete disappearance of compensation (*Q*-type *M-T* curve).

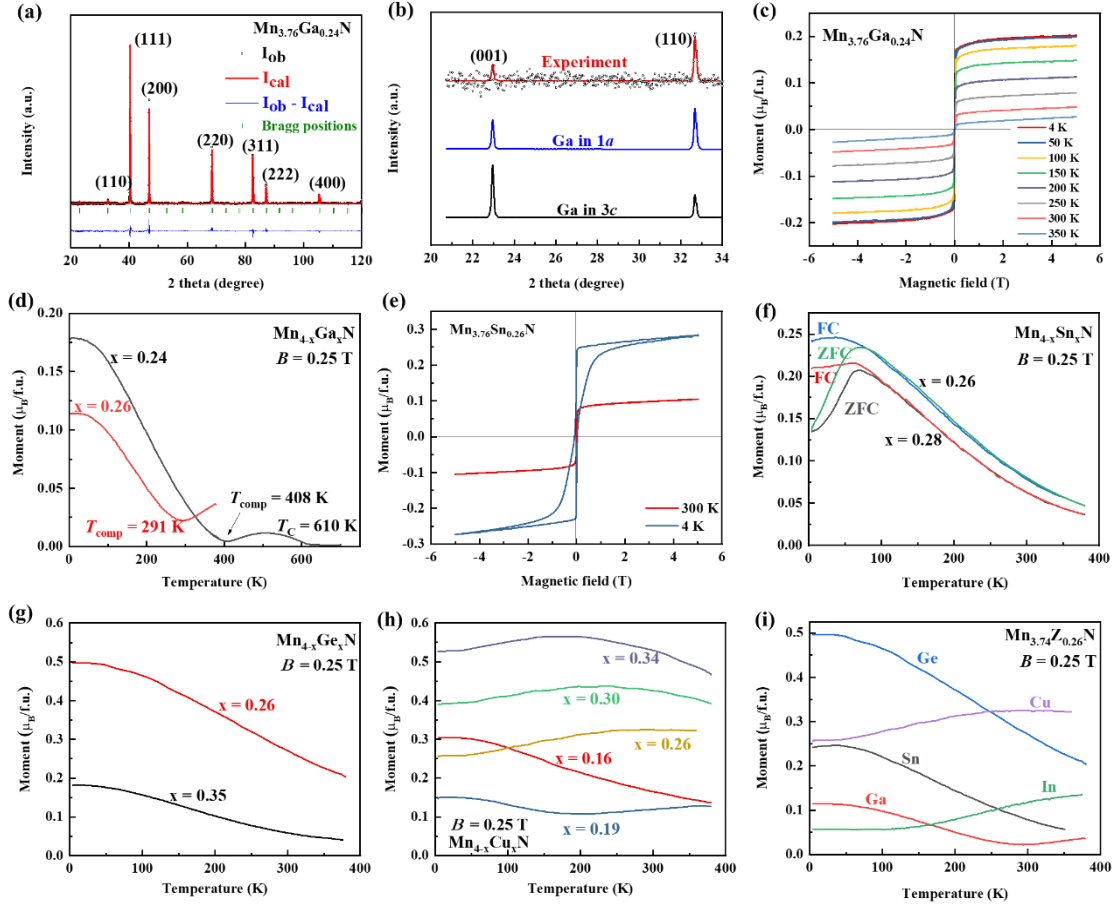

Fig. 3. (a) XRD pattern of $Mn_{3.76}Ga_{0.24}N$. (b) Expanded low-angle data with simulations showing the superlattice peak for $Mn_{3.76}Ga_{0.24}N$. The experimental data confirm that Ga atoms occupy the $1a$ site. (c) Magnetization curves for $Mn_{3.76}Ga_{0.24}N$ at different temperatures. (d) Thermomagnetic scans $Mn_{4-x}Ga_xN$ (x = 0.24 and 0.26). (e) Magnetization curve for $Mn_{3.74}Sn_{0.26}N$ at 4 K and 300 K. (f) Zero-field-cooled (ZFC) and field-cooled (FC) thermomagnetic scans for $Mn_{4-x}Sn_xN$. (g) Thermomagnetic scans for $Mn_{4-x}Ge_xN$ (x = 0.26 and 0.35). (h) Thermomagnetic scans for $Mn_{4-x}Cu_xN$ (x = 0.16 to 0.34). (i) Thermomagnetic scans for $Mn_{3.74}Z_{0.26}N$ (Z = Cu, Ga, Ge, In and Sn).

We plot data for different compositions of $Mn_{4-x}Z_xN$ at 4 K in Fig. 4a) including both our own data and previous reports [26,30,31,32]. The slope changes significantly from Ni (-6.20 $\mu_B$/atom), Cu (-5.01 $\mu_B$/atom), Zn (-4.35 $\mu_B$/atom), Ga (-3.70 $\mu_B$/atom) to Ge (-2.52 $\mu_B$/atom) with increasing valence electron for dopants in the fourth period. A similar trend is also found for dopants in the fifth period: Ag (-7.70 $\mu_B$/atom), In (-4.35 $\mu_B$/atom) and Sn (-3.08 $\mu_B$/atom). Based on this trend, the magnetic diagrams for different types of *M-T* curve are visualized in the maps of Figs. 4c and 4d. With a small concentrations of dopants, the interaction between $1a$ and $3c$ sites dominates and there is no compensation below the Curie temperature leading to a *Q*-type *M-T* curve. With suitable *x*, the moment if the $1a$ sublattice is still larger than that of the $3c$ sublattice at low temperature, while at high temperature the $3c$ sublattice wins above compensation. Therefore, an *N*-type *M-T* curve is found. When heavily-doped, the $3c$ sublattice dominates throughout whole temperature range, and there is a *P*-type *M-T* curve with no compensation. Elements from the fifth period have a greater ability to compensate than those from the fourth period, and the magnetization is very sensitive to *x*. Therefore, the boundary for different *M-T* curves are shifted to the left (lower *x*) and the useful *N*-type region is narrower.

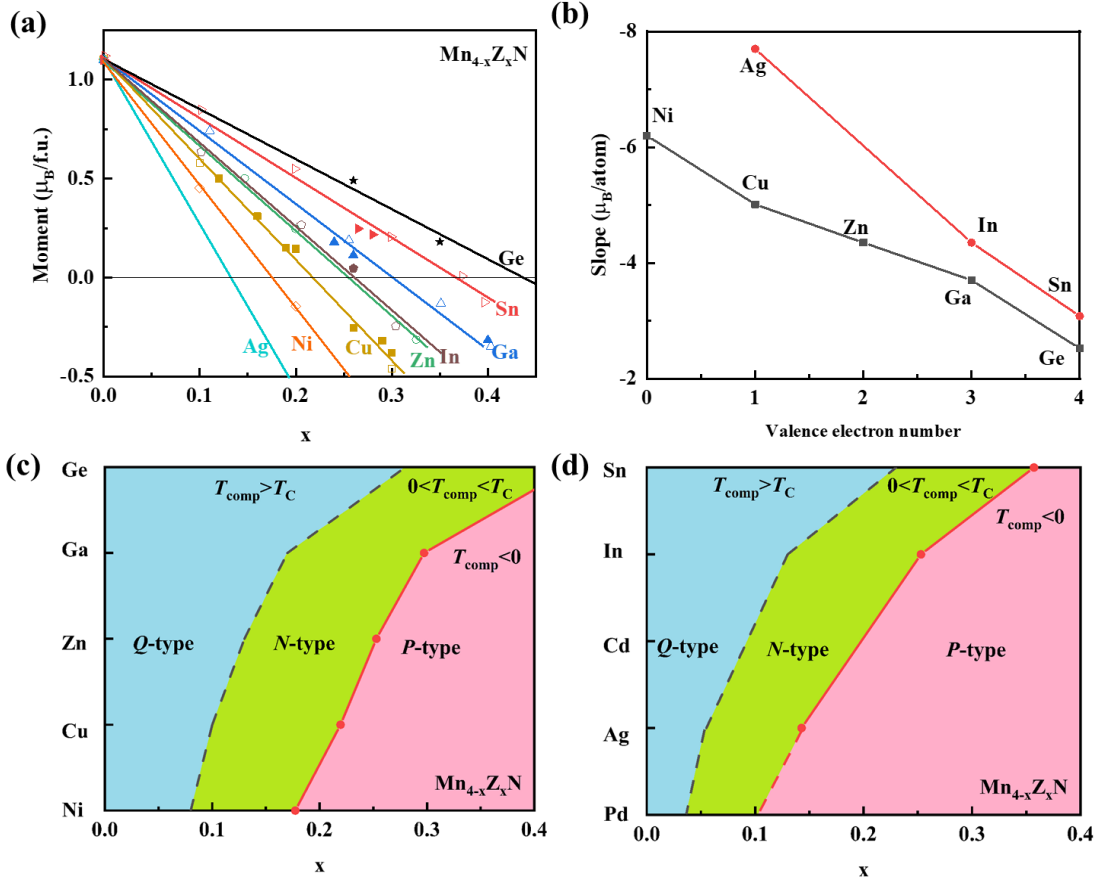

Fig. 4. (a) Summary of the net moment as a function of composition x in $Mn_{4-x}Z_xN$. We include our own data (solid points) and previous reports (open points). (b) The slope in (a) showing different efficiencies. Magnetic diagram for *Q*, *N* and *P* type *M-T* curves, depending both on x and Z for elements from the fourth (c) and fifth (d) periods.

## 4. Discussion

The efficiency of the dopants to compensate is analyzed from the view of magnetic moment, non-collinear angle and lattice constant, from both experimental and theoretical points of view.

### 4.1 Lattice constant

Compounds with the same *x* but different Z from the same group have the same valence electron number, and the main difference in their effects on the magnetic structure is related to the lattice parameter. Fig. 5a shows $a_0$ for $Mn_{4-x}Z_xN$ (Z = Cu, Ga, Ge, Ag, Sn In). It is clear that Ag [33], In and Sn [26] lead to a greater increase in lattice parameter than Cu, Ga and Ge for the same *x*, because of their larger atomic radii. The increased lattice parameter translates to a larger atomic separation in the cubic crystal, leading to reduced *p-d* hybridisation of $Mn^{3c}$ and increased Mn-Mn exchange that produce a relative increase of magnetism of the 3*c* sublattice. The $Mn^{3c}$ moment is larger and more localized, leading to improved doping efficiency for dopants from the 5[th] period.

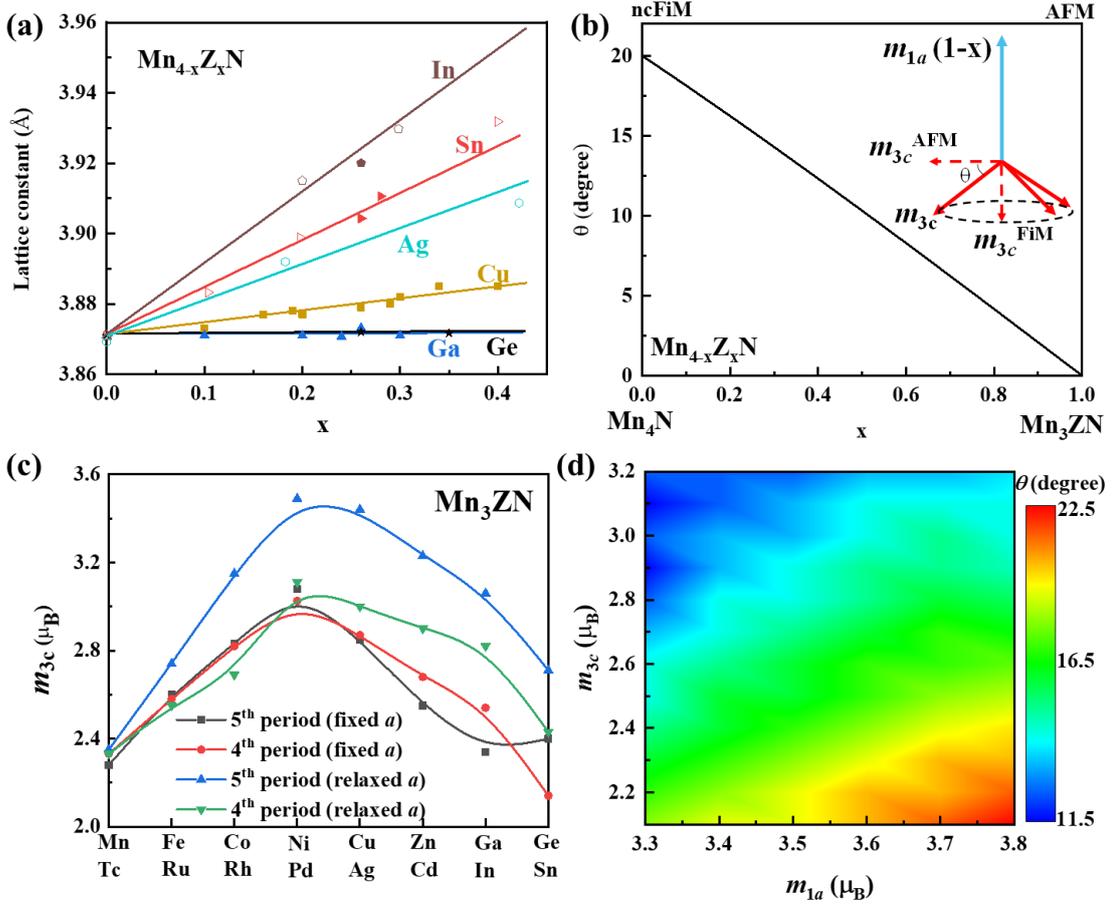

Fig. 5. (a) Lattice parameters for $Mn_{4-x}Z_xN$ (Z = Cu, Ga, Ge, Ag, Sn In), comparing our data (solid points) and previous reports (open points) [26,33]. (b) Calculated tilt angle $\theta$ versus x by Eq. 6. (c) Calculated magnetic moment for $Mn^{3c}$ in $Mn_3ZN$ ($x = 1$). (d) Tilt angle $\theta$ for different $m_{1a}$ and $m_{3c}$ deduced from DFT.

**4.2 Magnetic moment**

Since the nearest-neighbor for $Mn^{1a}$ is always $Mn^{3c}$, the magnetic coupling between $Mn^{1a}$ atoms is weak. As a result, the moment for the remaining $Mn^{1a}$ is not influenced significantly, as also indicated from neutron diffraction [34]. Therefore the effect of doping on the net magnetization comes mainly from $Mn^{3c}$. We built our DFT model to capture trends in the electronic and magnetic structures and used the simplest 5-atom unit cell model to model our experimental observation with different compositions. This model allows us to compare trends for $x = 1$, when the 3c site is fully occupied by Mn and the symmetry is cubic.

The calculated $Mn^{3c}$ moment as a function of valence electron count in $Mn_3ZN$ from Mn to Ge in the 4[th] period and from Tc to Sn in 5[th] period is plotted in Fig. 5c. The magnetic behaviour shows the same trend; an initial increase saturates around Ni and Pd, and drops monotonically afterwards. In order to separate the electronic effects from the impact of chemical pressure on the lattice parameter, we first fixed $a_0$ of all members of the series to that of $Mn_4N$ (3.82 Å). This is shown by the solid red and black lines for the 4[th] and 5[th] periods in Fig. 5c. The peak at Ni, which has three electrons more than Mn, resembles a localized moment picture with striking similarities with the Slater-Pauling rule. These three extra electrons are shared by the three nearby Mn$^{3c}$ atoms, and hence each of the Mn$^{3c}$ atoms get one more electron becoming like iron, which shows the largest average moment in 3$d$ alloys. The influence of the lattice constant on the moment expected on Mn$^{3c}$ without constraint is also drawn in green and blue lines for comparison. The same trend is maintained, with peaks at Z = Ni and Pd. The main difference is found on the right hand side of the curves, especially for elements from 5$^{th}$ period, as it relates to the expanded lattice parameters with additional valence electrons compared to the Z = Mn reference. The significant change in the amplitude of $m_{3c}$ is one of the main reasons for the different doping efficiencies. In addition, the orientation of $m_{3c}$ also depends on its amplitude that further impacts the efficiency of compensation as we discuss it in the following section.

**4.3 Tilt angle**

Two questions concerning the tilt angle $\theta$ between $\boldsymbol{m}_{3c}$ and the (111) plane are: How does $\theta$ change with x, and with the different dopings?

The 3$c$ moment has two components, one component $m_{3c}^{Fi}$ along the ferrimagnetic [111] axis and the other $m_{3c}^{AFM}$ in triangular antiferromagnetic (111) plane. The molecular field acting on 3$c$ site also has two components, parallel and perpendicular to the [111] axis $H^{Fi}$ and $H^{AFM}$, corresponding to the ferrimagnetism and in-plane antiferromagnetism. They satisfy the relationships

$$H^{AFM} = -2\, n_{3c3c}\, m_{3c} \cos\theta \cos120° \quad (1)$$
$$H^{FiM} = n_{1a3c}\, m_{1a}(1 - x) - 2n_{3c3c}\, m_{3c} \sin\theta \quad (2)$$
$$m_{3c}^{Fi} = m_{3c} \sin\theta \quad (3)$$
$$m_{3c}^{AFM} = m_{3c} \cos\theta \quad (4)$$
$$\tan\theta = H^{FiM} / H^{AFM} \quad (5)$$

where $n_{3c3c}$ and $n_{1a3c}$ are the Weiss coefficients for interactions between 3$c$-3$c$ Mn and 1$a$-3$c$ Mn, as shown from the inset of Fig. 5b. In Eqs. 1 and 2, the in-plane antiferromagnetism considers the interaction from the other two nearest neighbor Mn$^{3c}$ with 120° triangular spin structure; the negative sign of $m_{3c}$ is already considered. Taking Eqs. 1 and 2 into Eq. 5, we get

$$\sin\theta = n_{1a3c}\, m_{1a}(1 - x)/(3n_{3c3c}\, m_{3c}) \quad (6)$$

Previously $\theta$ was estimated from neutron diffraction to be about 70° (nearly collinear), with $m_{3c}^{Fi} = 0.9\ \mu_B$, $m_{3c}^{AFM} = 0.36\ \mu_B$, $m_{3c} = 0.97\ \mu_B$, $m_{1a} = 3.8\ \mu_B$ but with large error bars [13]. This means the doping efficiency should be weaker than –(3.8+0.36) = -4.16 $\mu_B$/atom, if we assume that the magnetic structure becomes collinear ferrimagnetic after doping. But from both our and previous experiments, dopants of Ni, Ag, Cu are much more effective, indicating that $m_{3c}^{AFM}$ was underestimated. This is also found in our DFT calculations, where $\theta = 19.5°$ for binary Mn$_4$N. Thus if we plot the relationship between $x$ and $\theta$ in Eq. 6 (Fig. 5b), we find that $\theta$ decreases with $x$ almost linearly. The umbrella-like triangular spin structure of Mn$^{3c}$ rotates away from the [111] direction and becomes in-plane and therefore the net moment changes at a slower rate, as confirmed by comparative neutron study of Mn$_{3.2}$Ga$_{0.8}$N and Mn$_4$N [34]. Finally, when x = 1, Mn$^{1a}$ is completely replaced by the nonmagnetic dopant, Mn$_3$ZN is a triangular topological antiferromagnet in the (111) plane if the crystal remains cubic.

When doped with different elements from Cu to Ge, or from Ag to Sn, the decrease of $m_{3c}$ with increase of valance electrons leads to a rise of $\theta$ according to Eq. 6. This can weaken the effect of decreasing $m_{3c}^{Fi}$ according to Eq. 3. Similarly, considering the increased lattice constant for dopants from 5$^{th}$ period, the enhanced $m_{3c}$ can also lead to a drop of $\theta$, weakening the influence on $m_{3c}^{Fi}$. We further estimate $\theta$ by DFT calculation based on different $m_{1a}$ and $m_{3c}$ manganese site moments, as shown in Fig. 5d. We use a fixed spin moment (FSM) approach, where the amplitude of the magnetic moments on both sites is fixed. In Fig 5d, we plot the angles for minimum total energy $E_{tot}(m_{1a}, m_{3c})$. The general trend is that the larger $m_{3c}$ for a given $m_{1a}$, the smaller $\theta$. The larger $m_{3c}$ moment tends to stay in the (111) plane, and only the increasing moment on the 1a site could compensate for this rotation. This is in qualitative agreement with Eq. 6 and the vanishing moment on $m_{1a}$ with increasing $x$, from experiment.

**4.4 Best dopants for compensated ferrimagnetism**

$Mn_{4-x}Z_xN$ thin films are already attracting increasing attention for spintronics [35,36]. Most studies have been done with Ni or Co [22,23,24,25], but they are not ideal for achieving compensation. Beside the demand for compensation, additional requirements must be considered when choosing the best dopants. Based on our analysis, Ga appears to be a suitable dopant in $Mn_{4-x}Z_xN$ films for spintronics for the following reasons: First, earlier elements from 4$^{th}$ period like Ni compensate the moment with small values of $x$. As the total moment is very sensitive to the composition, it is difficult to control the composition precisely and homogeneously. Second, for elements that have many additional valance electrons like Ge, a large value of $x$ is needed due to the low doping efficiency. As a result, the Curie temperature drops substantially, which is not beneficial for room temperature applications. Third, Ga does not significantly increase the lattice constant compared to elements from 5$^{th}$ period so that a series of thin films can be grown with different compositions $x$ and a similar tetragonal distortion is expected on the same substrate. The slight tetragonal distortion ($c/a \sim 0.99$) due to biaxial strain imposed at the interface of the film and substrate is the origin of perpendicular [001] anisotropy. A smaller lattice constant of the film than common substrates, such as $SrTiO_3$ with $a_{001} = 3.91$ Å, is the key for the in-plane tensile strain and perpendicular anisotropy [37,38], which can be easily realized in Ga-doped samples. Finally, the doping efficiency of Ga, -3.70 $\mu_B$/atom coincides with $m_{1a}$. This is a consequence of a combination of an increased $m_{3c}$ and a decreased $\theta$ rather than simply the nonmagnetic nature of the dopant.

**5. Conclusion**

From our experimental and theoretical study of the rare-earth-free noncollinear ferrimagnetic metals $Mn_{4-x}Z_xN$, we conclude that the noncollinear ferrimagnetism originates from the structure of the $Mn^{3c}$ (111) kagome planes with a small Mn-Mn interatomic separation that leads to frustration of the antiferromagnetic nearest-neighbor interactions. The tilt angle of the moments from the (111) planes, $\theta = 20°$, is smaller than previously thought. There is a choice of substitutions to achieve magnetic compensation at room temperature. The efficiency of different elements in this respect rises gradually with increasing valance electrons from group 11 (Cu, Ag) to group 14 (Ge, Sn). The $Mn^{1a}$ moment is not sensitive to the dopants, while the $Mn^{3c}$ moment peaks at Ni and Pd then drops with further valance electron addition. The higher efficiency of elements from 5$^{th}$ period is due to the larger lattice constant, which weakens the hybridization of $Mn^{3c}$ and N and leads to an increase of $m_{3c}$. In addition, the tilt angle decreases with increasing $m_{3c}$ and composition $x$. Based on the above results, Ga is the recommended dopant for $Mn_{4-x}Z_xN$ for thin film spintronics with perpendicular anisotropy.

**Acknowledgement**

Yangkun He and Rui Zhang contribute equally to the work. We thank Yu Kang for high-temperature measurements. This work was supported by Science Foundation Ireland, under the MANIAC, SFI-NSF China project (17/NSFC/5294) and ZEMS project 16/IA/4534.